Demandance

Jeff Shrager (jshrager@stanford.edu)
Senior Fellow, CommerceNet
and Consulting Professor, Symbolic Systems Program, Stanford University

ABSTRACT

A demandance is a psychological "pull" exerted by a stimulus. It is closely related to the theory of "affordance". I introduce the theory of demandance, offer some motivating examples, briefly explore its psychological basis, and examine some implications of the theory. I exemplify some of the positive and negative implications of demandances for design, with special attention to young children and the design of educational products and practices. I suggest that demandance offers an approach to one of the persistent mysteries of the theory of affordance, specifically: Given that there may be many affordances in any particular setting, how do we choose which to actually act upon?

FROM AFFORDANCE TO DEMANDANCE

At about 50 minutes into Disney's "Who Framed Roger Rabbit", Judge Doom, the villain, is trying to extract our hero, Roger Rabbit from his hiding place in a room behind a bar. Roger is a cartoon, and Judge Doom knows that "No Toon can't resist the old 'Shave and a Haircut'," so he begins tapping out the first five beats: "Shave and a haircut…" on various walls and objects in the bar with his cane, pausing precipitously for a moment after each iteration. Indeed, Roger is unable to resist. He begins shaking uncontrollably, and finally bursts through the wall to complete the phrase in grand style: "Two bits!" This scene brilliantly demonstrates the phenomenon that I refer to as "demandance", a psychological urge exerted by a stimulus. Demandance is closely related to the concept of "affordance", an important concept in product design and analysis (Norman, 1988). The term "affordance" was adopted by Don Norman from J. J. Gibson's ecological psychology (Gibson, 1979). For Gibson, affordances were (roughly) latent functional possibilities in a particular setting, often although not always, "directly perceived" by an organism. For example, a flat horizontal surface "affords" landing on for a bird. Norman adapted this concept to the analysis of product and interface design, and there has been a great deal of exploration and argument regarding the concept (e.g., Norman, 1999; McGrenere & Ho, 2000; Oliver, 2005).

Whereas affordances "afford", or offer actions, demandances *demand* them. All people uncontrollably orient to bright lights or loud sounds. Infants are unable to resist sticking things into their mouths, and toddlers are the same way with candies. And, more subtly, just as Roger Rabbit was unable to resist Judge

Doom's entreaty, we expect the last note in a musical scale or familiar musical phrase, and mentally complete it when it isn't completed by the musician. In some such cases we have explicitly recognized and respected this phenomenon. For example, engineers have created child-proof caps in response to the problem of children tending to want to eat things; especially drugs, which look like candy. However, cognitive scientists do not have a name for this phenomenon, nor, more importantly, much in the way of analysis.

Certainly children are an important domain for design, and I hypothesize that demandances are at least as important in the design and analysis of products intended for young children as are affordances. The age range of concern here is primarily about 6 months through 3-4 years; just about from the time a child can execute independent complex action sequences, to the time that they have enough supervisory control over these action sequences to be able to suppress the pull exerted by demandances. However, demandance is not only a childhood phenomenon, and I later will provide examples from adult activity and design as well.

In what follows I describe some details of the concept of demandance, including a brief exploration of the psychology of the phenomenon. I also describe both its positive and negative implications for design. I conclude by suggesting that demandance may offer an approach to one of the persistent mysteries of the theory of affordance, specifically: Given that there may be many affordances in an particular setting, how do people choose which one to actually act upon?

DEMANDANCE TO DISTRACTION

Demandance is a psychological "pull" (or "pressure") exerted by a stimulus – oftentimes to execute some specific action. These pulls can vary in strength; some people can stand up to some of them, and some cannot. Demandances are at least distracting, and sometimes dangerous, as in the case of child-safe caps or tweets arriving while we are driving.

Most engineered products have a primary function. For example, the primary function of a toy electronic piano may be to play, or learn to play, the instrument. Unfortunately, there is a tendency, especially in children's products, for designers to include secondary "side" functionalities, and, moreover, to decorate these side functionalities with distracting "push me" demandances, such as bright colors and flashing lights.

My favorite example of this is an electronic piano that I bought for my son when he was about two years old. It had perfectly passable keys and musical electronics, but was loaded with all manner of "push me" demandance-decorated side functionalities such as demos, background beats, flashing lights, and so on. Having side functionalities isn't per se bad, but side functionalities tied to demandances, such as flashing lights, can be quite a problem for those easily distracted, especially children. In a terrible case of a "demandance fail", this particular electronic piano that had a power button that flashed with the music when my son played anything, leading the child to press the power button, and thereby *turn off the piano every few notes*! I initially tried taping over the errant button, but then couldn't turn the piano on reliably. That device was in the trash within a week.

Of course, most readers of this paper have the willpower to suppress most of the demandances that they might come across. However, insidious demandances often creep into adult products as well. One commonplace problematic demandance is in smart-phone app ads, which, the screen space being so small, often show up in places that one wants to either look or click as a part of the natural workflow. As a result, these ads are often clicked by accident – or, in the present terms, they "demand" unintended clicking. Indeed, recent reports indicate that accidental clicking on phone app ads is quite common, possibly nearing half of all ad clicks on these devices (Paidcontent.org, 2011).)

Of course, the demandance quality of ads in phone app settings is much worse for children, and so-called educational and/or children's software that includes ads invariably leads to frustration for everyone involved as a result of the demandances leading to children clicking anything that moves. (We may need a new term, something like "demandentally" for an unintended, "accidental", action that was demanded by the design of the system.)

Even aside from ads, there are often problematic demandances in apps that are not so easy to resolve. For example, various sorts of secondary controls, such as those that return to menus or set preferences, often have problematic demandances, which, like the flashing power button on the piano, will quite often mislead my son into pushing them, even though he has found many many times that they lead him into dead ends (or worse – in-app purchases!) Indeed, I've created a set of cardboard overlays for my iPhone and iPad, which, when taped on, covers the parts of the screen that include the problematic controls. But this is a clunky work-around. The app designers could have simply made secondary controls that are less overtly demanding so that children do not execute them "on demand". For example, the designer would requiring the pressing of three corners of the screen at once, or some-such action that is difficult for a child to understand and/or do accidentally (demandentally?).

Most parents will explore toys intended for their children, and it is very common for toys targeted at a slightly older age group to be made available to the younger children by their siblings, resulting in frustration when good design is not maintained throughout the plausible age-range. One example of such thoughtful design would be hiding the on/off switch, say on the back of the product. This will not significantly bother older children, but will significantly help the younger ones avoid accidentally turning off the product. Indeed, the iPhone itself (aside from particular applications) is actually very well designed for children along these dimensions; most of the controls are non-obvious and relatively hard for small hands to press. The home button, which is much easier to press, is at least not flashy, even if it is right on the front of the device. It would be better if app designers could repurpose or disable the Home button, but Apple explicitly disallows this. Therefore, if the child does get it into his or her head that they can use the home button to escape from an app, there's nothing that the app designer can do about it. (My cardboard overlays cover the home button as well.)

The best example of an almost perfect application along the demandance lines is a product called "AlphaBaby" (Dickey). When the child types (or more commonly, pounds) on the keyboard, AlphaBaby displays letters, numbers, and shapes on the screen and makes "interesting" noises (if you're a 2 year old). Once AlphaBaby starts up, nothing that you can do, short of an OS-based "kill" will turn it off; every button or keyboard combination that the child is likely to press does something "interesting" (again, if you're a 2 year old). The one brilliant exception to this makes my point clearly: If you type the letters: q-u-i-t, AlphaBaby displays the q-u-i on the screen as normal, but then on the 't', it quits!

The astute reader, familiar with young children, will correctly point out that once a child figures finds the on/off button, or sees the obscure three-finger gesture, they will do it even if you hide the switch. This is certainly true if you make the on or off exciting, by, for example, giving it a startup sequence with exciting noises and flashing lights. But the theory of demandance contends that a demandance, in effect, inserts a goal into the person's mental set, so "out of sight out of mind" is the essential design principle for avoiding unintended demandances.

DEMANDANCES IN THE REAL WORLD

Demandance is not merely a product design issue. Problematic demandances play out quite often in the real world. An excellent example of this is the unfortunately common tendency of cyclists who are coming up from behind on an unsuspecting pedestrian, to shout something like "on your left", or merely just "left", meaning that the cyclist intends to pass the pedestrian on the latter's left. Now, whereas the cyclists surely have the best of intentions, let's consider this from the point of view of the pedestrian: You are walking along a narrow path, say with just enough room for two people to pass side by side. Someone suddenly shouts something at you from behind. Your automatic response, regardless of what they shout, is to jump out of the way, and the only place to jump is directly into the path of the on-rushing cyclist. To make matters worse, the cyclist yells "left", which you need to actually invert in order to move in the intended (safe) direction – i.e., to the right! Again, the most obvious thing to do when someone yells "left" at you is to move to the left. (The reader, especially those who are bicyclists, may doubt that this actually happens, but it has happened to me twice, once to the effect of minor scrapes, but once also resulting in a trip to the hospital with a fracture. Of course, in both cases the cyclists were angry for the person having moved right into their way, but it's not as though he had time to think about it – he was just responding to the demandance.)

The cyclist, being the only one who is aware of the impending collision, could defuse the demandance structure of this situation by announcing his or her approach much earlier so that the participants have time to problem-solve the situation safely, and not use confusing terminology. This used to be the function of a bike bell or horn. The complex implications of surprising people, and moreover shouting potentially confusing instructions at them, is elucidated by this analysis in terms of demandances.

Another, more positive, example of the use of a demandance is offered by extremely loud fire alarms, which not only wake up pretty much anyone, even with earplugs, but are often so loud that they demand that people vacate the area just to avoid the physical ear pain.

The foregoing analyses also point up the quasi-perceptual nature of some demandances, akin to Gibson's notion of the "direct perception" of affordances. Regardless, most design-related demandances, like most design-related affordances, are almost certainly learned; they depend upon the cognitive background of the user, as well as the user's interpretive stance; at least for adults, although perhaps less so for young children.

BALANCING POSITIVE AND NEGATIVE DEMANDANCE

As with most design problems, some tradeoff is required between desirable focusing of attention and undesirable distraction. Fire alarms are an example where demandances are a desirable feature, and similarly with flashy ads (at least from the standpoint of the ad designer). Some well-designed electronic musical instruments put the power button on the back or out of the way, and one can imagine other ways to reduce the demandance of this necessary functionality, for example by not lighting it, or maybe even lighting it only when the device is actually off, thus putting the demandance in the correct "orientation". In this sense, demandance may be related to stimulus-response-compatibility (Proctor & Reeve, 1990).

EMPLOYING DEMANDANCES IN EDUCATION

I have described demandance as mostly problematic for the design of products for young children. Most adults can suppress the psychology pull of an undesired demandance, except in rare cases such as the bicycle example, or positive cases such as fire alarms. Generally, as in the piano example, demandances tend to sidetrack intended educational processes by distracting the learner, or worse (e.g., turning off the device!). However, one can use demandance to one's advantage in education settings.

I taught my son to read more-or-less fluently by late in his third year. (Although the son mostly didn't understand what he was reading, because he didn't know what most of the words meant.) I of course can't say specifically which of the many things that I did with my son led to his somewhat precocious reading. However, for the present discussion, it is worth noting two methods that may have had some value, and which make direct use of demandances. The first method that I used, which is quite commonly recommend, was to point to words all the time when I read with my son (Grabmeier, 2012). Indeed, I would never read without pointing to the words that he was reading. My moving finger obviously demanded his attention, but this is a pretty simple example of attention capture. The second thing that I did, which I believe is less common, and harder to describe, but which was almost certainly quite important in my son's early reading abilities, was to directly employ what I will call the "Judge Doom" methodology. Specifically, I would read using over-emphasized voicing, but would stop short (when my son was younger) of the last word in a sentence, or (when he was somewhat older) of longer and longer words, regardless of their position in the sentence. There is a rise in tone that is a bit hard to describe that demands one's interlocutor to complete the sentence, sort of like the completion of a musical phrase, or the "Shave and a Haircut…." demandance employed by Judge Doom. I would get to the desired point (just before the end of the sentence, or just before a big word that I knew that my son would have a little trouble reading), and then stop with that rising intonation, and at the same time pointing to the next word, inviting the child – indeed demanding him – to read the word. The pull to complete a musical sequence, or my upon my son to finish my sentences involves neither attentional capture nor affordance in a simple way.

Put in slightly more technical educational terms, I was employing demandance to scaffold my son's reading (Wood, Bruner, & Ross, 1976), that is, to support his reading just beyond his ability, thereby permitting him to complete only those elements that are within his range of competence, and at the same time stretching his range slightly on each encounter.

ASPECTS OF THE PSYCHOLOGY OF DEMANDANCE

Demandance is obviously related to concept of attentional capture (Folk, & Gibson, 2001), but, just as affordance goes beyond low level perception, demandance goes beyond attentional capture, and in similar ways, as the examples above demonstrate. The flashing light on the piano certainly captures my son's attention, but the capture of his attention does not in-and-of-itself demand that he reach over and press the button. Pressing a button is an affordance of the button, and a choice of the user, driven, in the present theory, by the demandance. Perhaps a demandance is a combination of attention capture, affordance, and goal-directed action, but examples such as the demandance that Judge Doom deployed against Roger Rabbit, or that I used to teach my son to read, suggest a more intricate and subtle mechanism.

Regarding the relationship between "built in" and learned demandances, as exemplified by the bicycle example, many demandances are probably based in "built in" or very early learned attention pulls in response to stimuli. There is a great deal of work in these areas, and I cannot do justice to it in this brief paper. However, there is some interesting and quite directly relevant psychological work that I think is useful in revealing unexpected nature of demandance. I have already mentioned Gibson's work on affordance. Another useful thread arises from the gestalt concepts of "closure", and the related "Zeigarnick" effect. Gestalt closure is usually taken to assert that partial percepts are automatically completed by our perceptual systems or, put the other way around, partial forms are perceived as incomplete and, in the present sense "demand" their own completion. The incompleteness, in and of itself, does not suggest a pressure to complete the form or phrase, but various closely related concepts do add this nuance, specifically the concept of musical tension (Paraskeva, & McAdams, 1997; Fredrickson, 1999), where there appears to be a measurable "pull" (described as "tension") to complete musical phrases, such as scales. Another is the Zeigarnick effect (Atkinson, 1953), wherein people remember incomplete goals, for example from interrupted tasks, better than completed goals. The Zeigarnick effect is usually stated in terms of high level goals (even though it arose historically from a gestalt tradition), and the memory probes in these experiments are usually delayed by minutes hours or days, whereas I intend demandances as closer to immediate. Indeed, the "shave and a haircut" phenomenon, so well employed by Judge Doom against Roger Rabbit, is a wonderful example that stands right between Gibsonian/Gestalt direct perception and more high level/longer term Zeigarnick effect.

Regarding the ability to suppress the psychological "pull" (or "pressure") of a demandance, there are a number of lines of research regarding "delay of gratification", especially in children (Mischel, Shoda, & Rodriguez, 1989) and a more general "self control" literature [14, note 18]. These experiments demonstrate the children around 4 years old are able to suppress demandances in some circumstances.

COULD DEMANDANCE REPALCE AFFORDANCE IN COGNTIVE THEORY

A persistent issue in the theory of affordances is this: If it is the case that a given setting generally affords many actions, how does the actor chose to which to respond.[1] In the SOAR theory of universal subgoaling (Laird, Rosenbloom, & Newell, 1986), having multiple otherwise undifferentiated possibly actions leads to the setting of a subgoal (and thus a subproblem) to figure out which one to do. But there are evidently situations where there are multiple possible action, but there is no apparent sub-problem being solved because the response time is essentially instantaneous. Indeed, it is almost always the case that there is more than one possible action – more than one available affordance – but actors are almost never lost in thought. One approach to resolving this conundrum is to suppose a "view" or "interpretive stance" (Shrager, 1987) which lifts certain affordances out from a background of all possible affordances. Thus, for example, when we approach an intersection we know, because we are familiar with driving and intersections, to attend to the other cars approaching from other directions, and that the car behind us generally does not matter, so we do not attend to it, nor generally do we respond to actions that it might afford. Put another way, we approach the intersection with, so to speak, an "intersection view", or "intersection blinders". Chisek (2007) offers a neurophysiological rendering of this theory.

---

[1] Thanks to Rob St. Amant for this insightful observation.

The concept of demandance offers a somewhat different way of thinking about action choice. Specifically, certain affordances have a stronger psychology pull or pressure than others, and those are the ones which we tend to choose. This can, of course, be over-ridden, especially by adults and especially in accord with a particular attentional set. But certainly demandances, such as headlights rapidly looming in the rear view mirror, are likely to cut through

CONCLUSION AND SOME RECOMMENDATIONS

Our goal in this paper was to introduce and exemplify the concept of "demandance", especially as relates to the design of products for young children, and especially educational products. To the extant that demandance is useful, it has both positive and negative implications for design of such products. I conclude a few vague recommendations regarding demandance-informed design:

1. Only make the central functions (and their affordances) available on the "focal" (most user-facing) interface. In the case of an electronic piano, for example, the central functions are those of the 88 keys. All others functions are secondary and should be off the main display.

2. If you cannot hide secondary functions at least do not decorate them, for example with flashing lights, and consider putting "dangerous" functions, like on/off switches under the moral equivalent of child-proof caps.

3. Similarly, as with child-proof caps, consider design with regard to the age range that the product is likely to reach, beyond the age range that you intend it to reach.

4. In educational settings take special note of distracting demandances, and employ demandances to scaffold learners.

ACKNOWLEDGMENTS

Leo Armel's years of impatient demandance led me to realize the importance of the concept.  Thanks also to Rob St. Amant, Pete Pirolli, Jack Carroll, and a number of other reviewers for their useful comments. Avery James carefully proofread the paper.